\documentclass{jfm}
\renewcommand{\figurename}{figure}
\usepackage[utf8x]{inputenc}
\usepackage{psfrag}
\usepackage{amssymb, amsmath}
\usepackage{siunitx}
\usepackage{color}

\usepackage{natbib}

\usepackage{url}

\DeclareSIUnit{\wtpercent}{wt\%}

\usepackage{upgreek}
\begin{document}

\shorttitle{Hydrodynamics of the phase-segregation in binary droplets} 
\shortauthor{Y. Li et al.} 

\title{Physiochemical hydrodynamics of the phase-segregation in an evaporating binary microdroplet}

\author{Yaxing Li\aff{1,2}
\corresp{\email{yaxili@ethz.ch}},
  Pengyu Lv\aff{3},
  Christian Diddens\aff{1}
 \and Detlef Lohse\aff{1,4}
\corresp{\email{d.lohse@utwente.nl}}
 }

\affiliation{\aff{1}Physics of Fluids group, Department of Science and Technology, Mesa+ Institute, Max Planck Center for Complex Fluid Dynamics and 
J. M. Burgers Centre for Fluid Dynamics, University of Twente, P.O. Box 217, 7500 AE Enschede, The Netherlands
\aff{2}Institute of Fluid Dynamics, Department of Mechanical and Process Engineering, ETH Z{\"u}rich, Switzerland
\aff{3}SKLTCS and Department of Mechanics and Engineering Science, BIC-ESAT, College of Engineering, Peking University, Beijing 100871, China
\aff{4}Max Planck Institute for Dynamics and Self-Organization, 37077 G\"ottingen, Germany
}

\maketitle

\begin{abstract}
Phase segregation triggered by selective evaporation can emerge in multicomponent systems, leading to complex physiochemical hydrodynamics. Recently, Li~{\it et al.}~({\it Phys. Rev. Lett.}, vol. 120, 2018, 224501) and Kim \& Stone~({\it J. Fluid Mech.}, vol. 850, 2018, 769) reported a segregative behavior (i.e., demixing) in an evaporating binary droplet. In this work, by means of experiments and theoretical analysis, we investigate the flow dynamics \textit{after the occurrence of the phase segregation}. As example, we take the 1,2-hexanediol-water binary droplet system. First, we experimentally reveal the overall physiochemical hydrodynamics of the evaporation process, including the segregative behavior and the resulting  flow structure close to the substrate. By quantifying the evolution of the radial flow, we identify three successive life stages of the evaporation process. At Stage I, a radially outward flow is observed. It is driven by the Marangoni effect. At the transition to Stage II, the radial flow partially reverses, starting from the contact line. This flow breaks the axial symmetry and remarkably is driven by the segregation itself. Finally at Stage III, the flow decays as the evaporation gradually ceases. At this stage the segregation has grown to the entire droplet, and the flow is again controlled by the Marangoni effect. The resulting Marangoni flow homogenizes the distribution of the entrapped volatile water over the whole droplet. 

\end{abstract}


\section{Introduction}\label{intro}

Evaporation of sessile droplets has been vastly studied in the past decades~\citep{cazabat2010,brutin2015,ZANG2019,lohse2020physicochemical,wang2022}. Beyond the scope of fundamental research, the interest also originates from its huge relevance to various technological and biological applications, e.g. inkjet printing~\citep{park2006,Lohse2022ARFM},  surface patterning~\citep{kuang2014}, disease diagnostics~\citep{brutin2011}, and microfabrication~\citep{kong2014}, among others. A comprehensive physical picture of the evaporative lifetime of small airborne droplets can even be crucial to better understand the spreading of the corona virus in the current pandemic~\citep{mittal2020,Chong2021,Wang2021PNAS}. 

While the evaporation (or analogous dissolution) behavior of pure droplets has been extensively well understood~\citep{deegan1997,popov2005,ristenpart2007,cazabat2010,gelderblom2011,stauber2014,lohse2015rmp,basu_2019,ZANG2019,chong2020,Cira2021}, much less research has been done on multicomponent droplets. However, most droplet systems encountered in nature and technology are multicomponent, containing either multiple solutes~\citep{bennacer_sefiane_2014,kim2016controlled,diddens2017evaporating,wang2021,williams2021,Wu2021,jiang2021} or dispersed particulates~\citep{science2003,Lauga2004,marin2011,lijun2021}. The difficulty stems from the complex physicochemical hydrodynamics brought by the selective evaporation of each component in the mixture systems, including the coupling of the mutual interactions between species~\citep{Brenn2007}, contact line dynamics~\citep{Burton2020}, shape formation~\citep{Pahlavan2021}, flow structures~\citep{christy2011,edwards2018,Li2019,did21a,moore2021}, and even phase segregation~\citep{Tan2016,Li2018,kim2018,karpitschka2018,colinet2020,li_jfm_2020}. Over the past 20 years, a growing interest in such complex evaporative systems appears in literature, which was recently reviewed by \cite{lohse2020physicochemical}.

In (partially) volatile multicomponent droplets, in general, phase segregation emerges, as such systems are out of equilibrium. This is due to evaporation~\citep{Tan2016,Li2018,kim2018,karpitschka2018,colinet2020} or dissolution~\citep{dietrich2017}, or crystallization~\citep{crystal2018,Li2020Langmuir}. The emerging segregative patterns commonly lead to a nonvolatile shielding or an entrapment, where one liquid on the surface of the droplets covers the others, hindering its further evaporation. A detailed understanding of the shielding and the entrapment is crucial for many industrial and technical applications which require precise control of the droplet drying time, e.g. inkjet-printed droplets drying on papers. Moreover, liquid-liquid phase segregations are also essential in many biochemical reactions~\citep{hyman2014}. As a very recent example, a resemblant arch-shaped liquid-liquid-phase-segregation pattern triggered by evaporation has been observed in an aqueous mixture droplet of polyethylene glycol (PEG) and dextran~\citep{Guo2021}, which is potentially a robust method for prebiotic compartmentalization. 

Our recent study on an evaporating 1,2-hexanediol-water binary droplet reported a liquid-liquid phase segregation of the non-volatile (at room condition) component, i.e., 1,2-hexanediol, during the evaporative process~\citep{Li2018}. We demonstrated that for that particular binary droplet system the concentration gradient on the droplet surface caused by the preferential evaporation of water only leads to a small gradient of the surface tension. The resulting solutal Marangoni flow is then initially insufficient to homogenize the component distribution within the entire droplet. Therefore the accumulated solute separates from the mixture, starting with the nucleation of a series of small 1,2-hexanediol droplets at the contact line. These then merge to arch-shaped structures. The droplet lifetime is significantly extended due to the shielding effect of this non-volatile component. The volatile water remains trapped in the interior of the droplet.

\begin{figure}
\centering
  \includegraphics[width=1\textwidth]{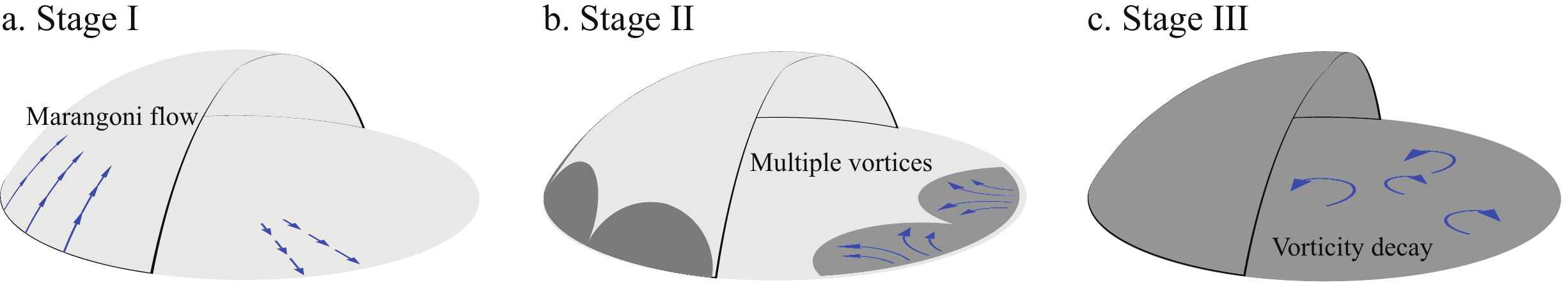}
  \caption{Schematics of the three life stages of an evaporating 1,2-hexanediol-water binary droplet.}
  \label{fgr:three_stages}
\end{figure}

In that study, we also showed that the flow structure drastically changes due to the emergence of the segregation. An evolution of flow structure was characterized by the temporal evolution of the vorticity (see Fig.~\ref{fgr:three_stages}):
\begin{enumerate}
  \item At the early stage before the segregation occurs, a radial flow near the substrate towards the droplet edge is driven by Marangoni forces.
  \item After the first emergence of a segregated phase of 1,2-hexanediol, the radial flow evolves to a state of multiple vortices.
   \item Eventually the vortices decay as the evaporation ceases.
\end{enumerate}
While the flow transitions were characterized by the evolution of vorticity, to the best of our knowledge, a comprehensive understanding of the complex flow dynamics induced by the segregation is still missing, especially the one within the segregative phase. Hence a systematic investigation is required. In this work, we thus revisit the evaporating droplet system of a 1,2-hexanediol-water binary solution with multiple experimental means, mainly focusing on the dynamics \textit{after the segregation has emerged}, aiming to systematically study the growth of the liquid-liquid-phase-segregation patterns and the resulting flow structure. We also offer a theoretical understanding of the local flow structure within the segregative patterns and of the decay of the flow intensity at the final stage of the evaporation.

The paper is organized as follows: In \S 2, we introduce the employed experimental techniques. In \S 3, we show the experimental observations on the overall physiochemical hydrodynamic of the evaporating process, and identify three successive life stages (Stage~I, II \& III) by the combination of the direct observation of the flow structures and the quantification of the temporal radial flow velocity. In \S4, we briefly show and discuss the flow structure at Stage I prior to the occurrence of the segregation, summarizing and confirming the results of \cite{Li2018}. In \S 5, we then focus on Stage II, in which the segregative patterns emerge and grow. We quantitatively interpret the growth dynamics of the segregative patterns and the local flow structures within them. In \S 6, we quantify and quantitatively explain the flow in the final Stage III, namely the decay of the flow intensity when the evaporation gradually ceases. The paper ends with a summary and further discussions (\S7).

\section{Experimental methods}\label{exp}
\subsection{Solution and substrate preparation}

The binary solution we used in our droplet system consisted of 90\%/w Milli-Q water (Reference A+, Merck Millipore, 25$^\circ$C) and 10\%/w 1,2-hexanediol (Sigma-Aldrich, $\geq$98\%). The experiments were performed on a hydrophobized glass slide coated by octadecyltrichlorosilane (OTS, Sigma-Aldrich, $>$ 90 \%)~\citep{Peng2014}.  Before each measurement, the substrates were processed with a sonication cleaning in 99.8 \% ethanol and in Milli-Q water for 15 min and 5 min sequentially, and subsequently dried with compressed N$_2$ gas flow for 30 s. The droplets were deposited by a glass syringe with a full metal needle (Hamilton, 10 $\mu$L, Model 701 NWG SYR, Cemented NDL).

\subsection{Confocal microscopic measurement}

Confocal microscopy was employed to conduct a series of measurements on the evaporation dynamics of the droplet. The observations were carried out by utilizing an inverted Nikon A1 confocal laser scanning microscope system (Nikon Corporation, Tokyo, Japan) with a 10x dry objective (Nikon, Plan Fluor 10x/0.30, OFN25, DIC, L/N1).

\subsubsection{Micro-particle image velocimetry}

For flow field visualization, we performed Micro-Particle Image Velocimetry ($\mu$PIV) by adding 
fluorescent particles [microParticles GmbH; PS-FluoRed-5.0: Ex/Em 530 nm/607 nm] into the
working fluids. The PIV measurements were also conducted on the confocal microscopy setup. The particles were excited by a laser at a wave length of 561 nm and the fluorescent signals were captured at a frame rate of 25 frames per second (f.p.s.). We measured the flow field near the substrate quantitatively by adding 520 nm diameter fluorescent particles at a concentration of $2 \times 10^{-3}$ vol\%
into the droplet. For prepossessing the PIV images, we removed the background noise outside the droplet by applying a contact line detection algorithm, and subsequently a pixel-wise minimum intensity background was subtracted from every image to enhance the signal to noise ratio. To obtain the velocity vector field, we performed iterative two-dimensional cross-correlations with multiple interrogation window sizes with a 50~\% overlap, which were 64 $\times$ 64 pixels, 32 $\times$ 32 pixels, and 16 $\times$ 16 pixels sequentially. A Gaussian fitting function was used to determine the subpixel displacement. The velocity vector calculation was performed using an open-source software, PIVlab~\citep{Pivlab2014}. 

\subsubsection{Refracted shadowgraphy}

We modified a method originally proposed by~\cite{chao2001}, namely laser refracted shadowgraphy, and applied it to the confocal microscopy. A white-light beam was collimated to a vertically parallel beam and then passed perpendicularly through a test droplet, placed on an OTS glass substrate. The white-light beam produced a
refractive image of the sessile droplet which was captured by the transmitted light detector at 25 f.p.s~\citep{lv2017}. The disturbances caused by local concentration variances within the test droplet refracted light rays, leading to shadows on the refractive images.

\subsubsection{Fluorescence image acquisition}
To visualize the local details within the segregative patterns, the binary mixture was labeled with two different dyes, i.e., Dextran and Nile Red~\citep{Li2018}. Dextran preferentially dissolves in water and it was excited by a laser at a wavelength of 488 nm, while Nile Red is a lipophilic dye that dissolves only in 1,2-hexanediol and was excited by a laser at a wavelength of 561 nm simultaneously. Two-dimensional (2D) images were obtained by scanning in the focal plane with a depth of field of 0.957 $\mu$m at the bottom of the droplet $\sim$10 $\mu$m above the substrate. It allows us to monitor the evolution of both the segregative patterns and the flow structure during the entire evaporation process. The scan started as soon as the droplet was deposited on the glass substrate. Operating in resonant mode, the 2D images were captured with a frame rate of 25 f.p.s.. To increase the contrast of images for further analysis, we converted the images to be binary (black and white) in colored.

\section{Overall phenomenology: Hydrodynamics induced by phase-segregation and the three identified life stages}\label{overall}

To obtain the overall phenomenology of the physiochemical hydrodynamics during the evaporation process in the 1,2-hexanediol-water binary droplet system, we first optically monitor the whole evaporation process, by using refracted shadowgraphy in combination with $\mu$PIV. The multiple channels of the confocal microscope allow us to perform the two visualization methods simultaneously, see Fig.~\ref{fgr:surface_flow}(a0-a2): 
\begin{enumerate}
   \item The transmitted light detector captures the refracted light through the droplet for shadowgraphy;
   \item The laser channel captures the fluorescent signal for $\mu$PIV.
\end{enumerate}
The focal plane is located at the bottom of the droplet slightly above the substrate ($\sim10~\mu$m). Fig.~\ref{fgr:surface_flow}(a1) displays the shadowgraph, in which the strong shadows indicate the segregative patterns, whereas the bright areas represent water-rich regions. Fig.~\ref{fgr:surface_flow}(a2) shows a typical snapshot of fluorescent particles for $\mu$PIV  at the same instant.

\subsection{Axial symmetric breaking induced by inhomogeneous nucleation of 1,2-hexanediol droplets}

\begin{figure}
\centering
  \includegraphics[width=1\textwidth]{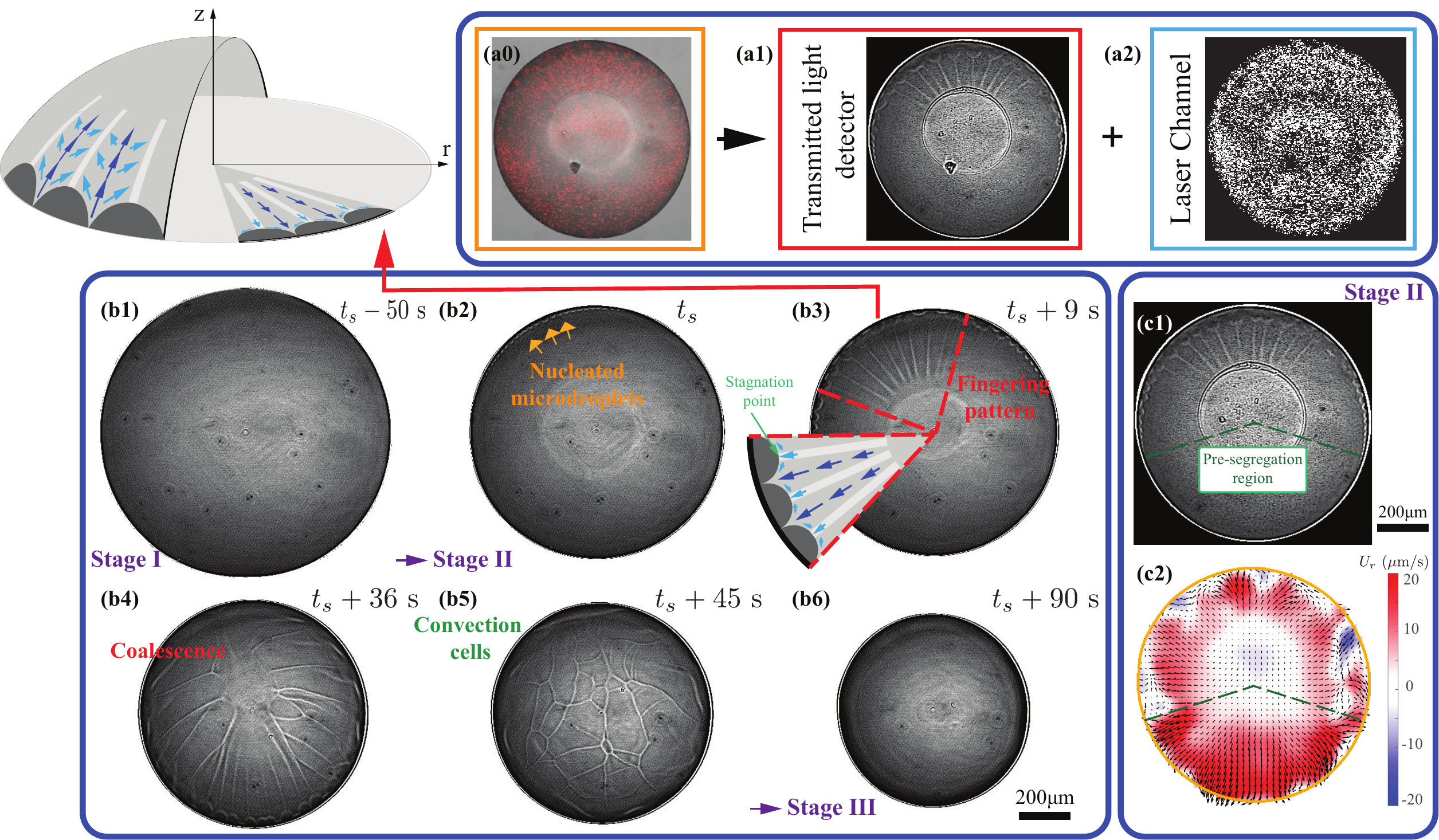}
  \caption{Confocal microscopic measurement of an evaporating 1,2-hexanediol-water droplet. (a) Transmitted light through the droplet and fluorescent signals of tracer particles simultaneously captured by two channels of the confocal microscopy. (b1-b6) Morphological evolution of convective structures with the emergence of phase segregation in an evaporating 1,2-hexanediol-water droplet. Here (b1) corresponds to Stage I, (b2), (b3), (b4) and (b5) to Stage II, and (b6) to the final stage III. The three-dimensional sketch illustrates the flow pattern in (b3). The dark blue arrows represent fast flow streams (inwards at the droplets surface and outwards close to the substrate) and the light blue ones  slow secondary flows. The white stripes represent the water-rich paths, which also correspond to slow-flow streams. The tip of each arch becomes a stagnation point of the slow-flow streams. (c) A refractive image of the droplet recorded by transmitted light detector (c1) and the corresponding flow field at the same instant (c2). Here the droplet is in Stage II.}
  \label{fgr:surface_flow}
\end{figure}

Fig.~\ref{fgr:surface_flow}(b1-b6) displays a series of refracted shadowgraphs of an evaporating 1,2-hexanediol-water droplet with the emergence of the segregation ($t = t_s = 50$~s). It starts with a homogeneous appearance of the entire droplet without the segregation at the beginning of the evaporation [Fig.~\ref{fgr:surface_flow}(b1)]. Then 1,2-hexanediol microdroplets nucleating at the rim of the binary droplet [Fig.~\ref{fgr:surface_flow}(b2)]. The emergence of nucleation sites is not homogeneously distributed along the contact line, but initially local at one certain point and only then extending to the rest of the rim. These inhomogeneous nucleations of segregative patterns induce an axial symmetric breaking of flow structures in the following process.

Without the pinning of the contact line, the contact area features a continuous shrinkage during the entire evaporation process. During the growth of the nucleated arch-shaped patterns, multiple water-rich streams emerge, revealed by the bright paths connecting between the center of the droplet and the tip of each arch pattern [Fig.~\ref{fgr:surface_flow}(b3)]. It resembles the thermocapillary instability appearing in heated droplets of ethanol~\citep{sefiane2008,saenz2017}. We note that the surface tension varies both with the local temperature and with the local solutal concentration, that is, $\gamma = \gamma_0 - \partial_T\gamma(T-T_0)-\partial_c\gamma(c-c_0)$; normally for binary droplets, $\partial_c\gamma(c-c_0) \gg \partial_T\gamma(T-T_0)$. Therefore the convective flow can be simplified to a purely soluto-capillary flow, neglecting the thermal Marangoni effect. 

The schematics in Fig.~\ref{fgr:surface_flow}(b3) illustrates a fingering flow pattern, namely the growing arches impede the Marangoni flow towards the edge of the droplet, causing the slow-flow stream. The tip of each arch becomes a stagnation point of the slow-flow streams (see sketch belonging to Fig.~\ref{fgr:surface_flow}(b3)). The radially outward flow in the streams continues towards the contact line region along the boundaries of the arches. During further evaporation, the arches grow and merge with neighbouring ones, which leads to the formation of new slow-liquid streams [Fig.~\ref{fgr:surface_flow}(b4)]. Due to the breaking of axial symmetry caused by the non-uniform growth of the segregative patterns, irregular convection cells appear and evolve [Fig.~\ref{fgr:surface_flow}(b5)]. Eventually, the convection cells disappear once the evaporation ceases [Fig.~\ref{fgr:surface_flow}(b6)]. As shown in the previous section, the expansion of the arches reverses the flow direction in the vicinity of contact line, namely from originally outward to inward. We qualitatively compare the shadowgraph with the $\mu$PIV results in Fig.~\ref{fgr:surface_flow}(c1-c2): the inward radial flow (indicated by blue color) in the contact line region shown in the velocity field [Fig.~\ref{fgr:surface_flow}(c2)] corresponds to the patterns within segregative region shown in the shadowgraph [Fig.~\ref{fgr:surface_flow}(c1)].

\subsection{Life stages identified by transitions in the flow}

To further quantify the evolution of flow structure, we analyze the measured results from the $\mu$PIV. In our previous study~\citep{Li2018}, the different life stages of an evaporating 1,2-hexanediol-water droplet were identified by the temporal evolution of the spatially-averaged vorticity. Here, by combining the $\mu$PIV results and the observations from shadowgraphy, more quantitative details of the flow transitions can be identified by the evolution of the local radial velocity. 

\begin{figure}
\centering
  \includegraphics[width=1\textwidth]{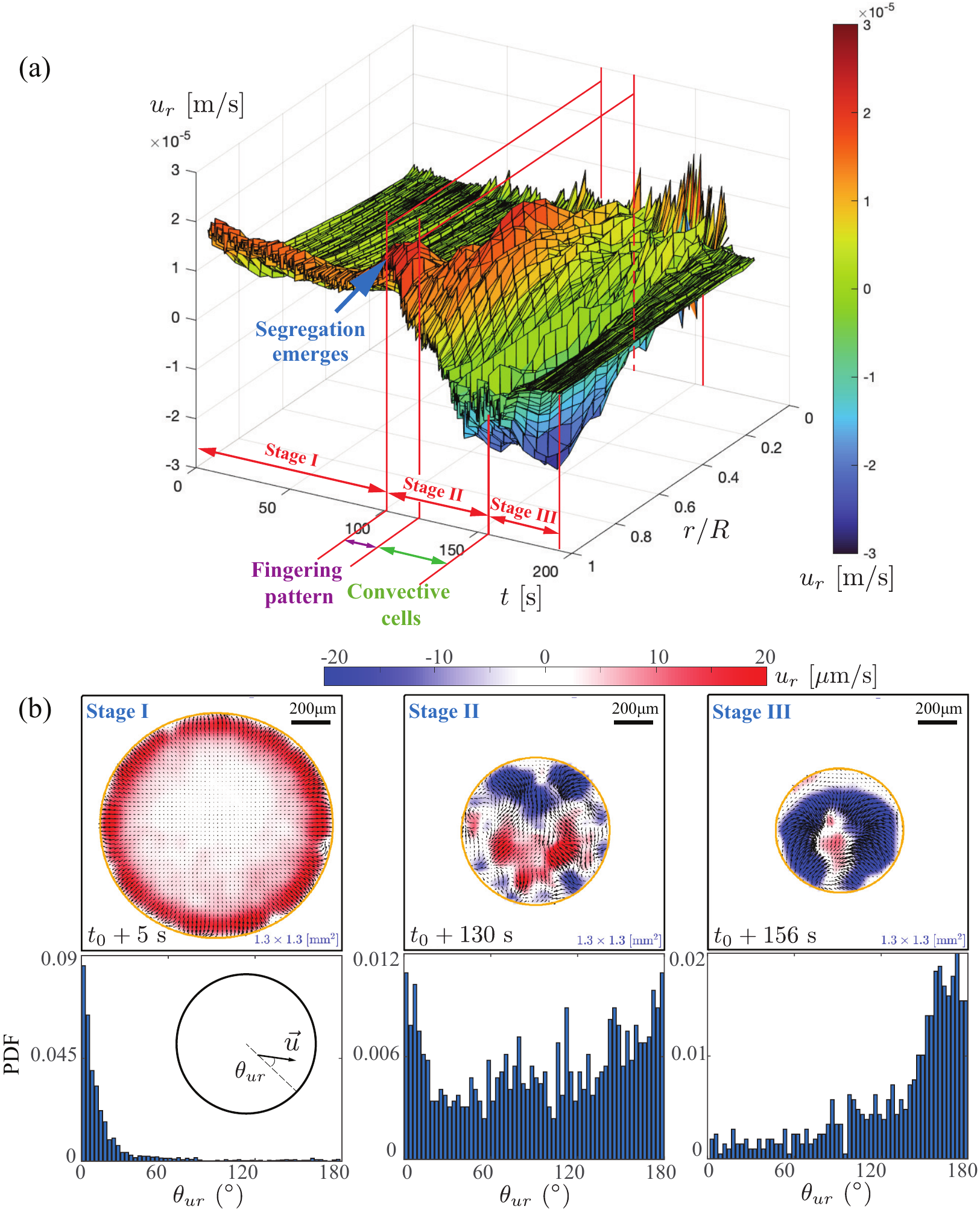}
  \caption{Results from micro-particle image velocimetry ($\mu$PIV) in an evaporating 1,2-hexanediol-water binary droplet. The flow field near the substrate is revealed. (a) Three-dimensional (3D) map showing the temporal evolution of the radial velocity averaged in azimuthal angle along the radial position. (b) The probability distribution function (PDF) of the azimuthal angle of the flow velocity vs. the radial direction at three different life stages.}
  \label{fgr:PIV-3}
\end{figure}

Fig.~\ref{fgr:PIV-3}(a) depicts a three-dimensional (3D) map showing the temporal evolution of the radial velocity along the radial position, which is normalized by the time-dependent footprint radius $R(t)$ of the droplet. The colour bar displays the intensity of the radial flow velocity, where positive values indicate outward direction. Three distinct stages of flow structure are identified:

\begin{enumerate}
  \item Stage I: Before the segregation occurs, the flow is radially outwards and most intensive in the vicinity of contact line, whereas the flow in the interior is barely visible;
  \item Stage II: With the emergence of segregation, the radial flow reverses from the contact line region, and the flow structure becomes irregular;
  \item Stage III: The flow intensity starts to decay to nearly zero, until the evaporation finally ceases.
\end{enumerate}
With the simultaneous observation from shadowgraphy, another two sub-regimes within Stage II can be distinguished, namely the fingering pattern and the convective cells shown in Fig.~\ref{fgr:surface_flow}(b3-b5). Stage III begins when the convection cells disappear. In this final stage the flow intensity starts to decay to nearly zero, until the evaporation finally ceases.

Fig.~\ref{fgr:PIV-3}(b) displays three snapshots of flow fields at different life stages. The first row shows the corresponding flow field at each instant. The second row shows the probability distribution function (PDF) of the azimuthal angle $\theta_{ur}$ of the local velocity (i.e., the angle between velocity $u$ \& radial direction) vs. the radial direction.The first column exhibits the flow structure in Stage I, which is before the occurrence of segregation. Driven by the Marangoni forces along the liquid-air interface, the flow is mainly radial towards the edge of the droplet (seen by the accumulation of PDF at $\theta_{ur} = 0$), where the flow is also more intensive than in the interior. With the emergence of segregation, Stage I undergoes a transition to Stage II. As seen in the second column, the PDF of the velocity direction in Stage II becomes more uniform, indicating a more irregular flow pattern, which can be seen in the snapshot of the velocity field. We attribute this to the existence of multiple vortices induced by segregation. The last column shows the velocity vector field at the beginning of Stage III, when the segregative patterns grow and finally fully occupy of the whole droplet. Then the PDF shows an opposite distribution as compared to Stage I, indicating that the inward flow radially towards the interior of the droplet is dominant. 

\section{Stage I: Marangoni-effect-dominant flow before segregation}\label{stagei}

In our previous work~\citep{Li2018}, we have studied the early period of the evaporation process before the occurrence of the segregation (i.e., Stage I), revealing the Marangoni-effect-dominant flow structure and the mechanism leading to the segregation. Due to the preferential evaporation of water, a relatively high concentration of 1,2-hexanediol is left behind and thus accumulates in the contact line region. This leads to a surface tension gradient at the liquid-air interface, driving a Marangoni flow from the contact line to the apex of the droplet. This interfacial flow also induces convective flows in the bulk of the droplet. Near the substrate, this convective flow is directed radially outwards to the rim of the droplet, which is consistent with our observation here in Fig.~\ref{fgr:surface_flow}(b1), and Fig.~\ref{fgr:PIV-3}(a). For details of this early Stage I and a quantitative analysis, we refer to \cite{Li2018}.

In the following sections of this paper, we focus on Stages II \& III and the segregative patterns and the resulting flow structures during these later stages, and offer theoretical interpretations of them.

\section{Stage II: Evolution of segregative patterns and induced local flow structure}\label{stageii}

At Stage II (after the first segregation has emerged), a radial flow reversal induced by the growth of the segregation occurs. To study this, a detailed local observation of the segregative patterns is required. Experimentally, we employ the fluorescence image technique coupled with confocal microscopy, i.e., labelling the segregated phase with different dyes, in order to locally resolve the details of the flow field within the segregated phase. Recently, \cite{Seyfert2021} reported an experimental study on the influence of added dye on Marangoni bursting phenomenon. In our case, the observations by labelling with fluorescent dyes show the emergence of more homogeneously distributed nucleation sites along the contact line, rather than the inhomogeneous nucleation sites there. Note that phenomenologically the two cases with or without the dyes show similar behaviors with respect to the growth of the segregation and the resulting flow structure within the segregative patterns. 

\subsection{Growth dynamics of the liquid-liquid phase segregation}
\begin{figure}
\centering
  \includegraphics[width=1\textwidth]{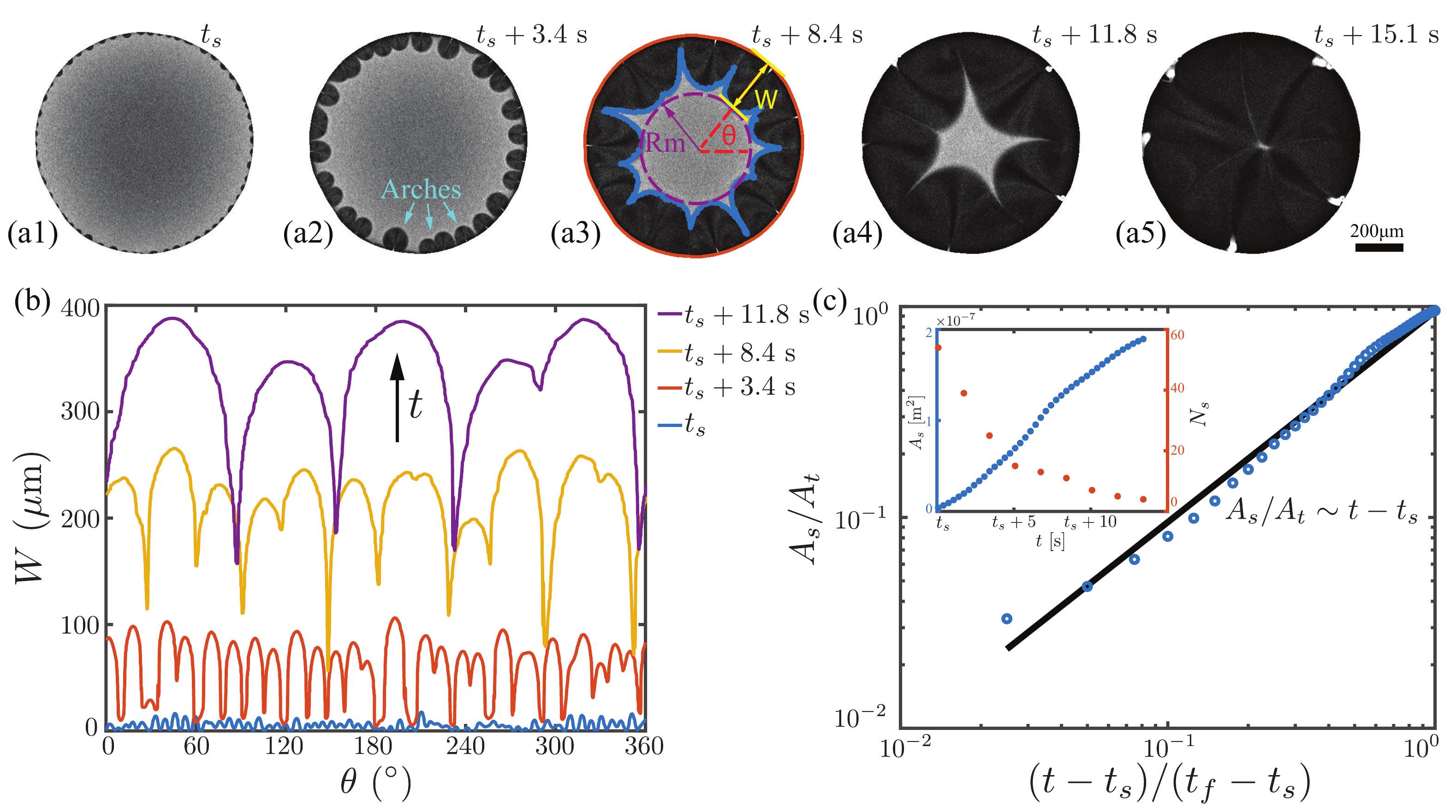}
  \caption{Growth dynamics of the segregated 1,2-hexanediol phase. (a) Bottom-view snapshots, captured via confocal microscope, highlighting the temporal growth of the phase segregation of 1,2-hexanediol in an evaporating binary droplet with 10\%/90\% (w/w) 1,2-hexanediol/water initial concentration. (b) Temporal evolution of the azimuthal undulations W($\theta$) of the nucleated droplets along the binary droplet radius during the growth of the segregated phase shown in sequence in (a). (c) Scaled master curve of the segregated phase area $A_s/A_t$ versus time $(t-t_s)/(t_f-t_s)$. Inset: Raw data for the temporal evolution of $A_s$ and the number of nucleated droplets $N_s$ versus the time $t$.}
  \label{fgr:nucleation}
\end{figure}

Before the study of the local flow structure within the segregative patterns of 1,2-hexanediol, we first aim to understand its growth dynamics. The images in Fig.~\ref{fgr:nucleation}(a1-a5) are taken by confocal microscopy from the bottom view: the dark regions shaping as arches in the vicinity of the contact line are the segregated phase of 1,2-hexanediol, whereas the grey counterpart is the mixture phase. The phase segregation starts with the emergence of various tiny nuclei evenly distributed along the contact line of the binary droplet [Fig.~\ref{fgr:nucleation}(a1)]. With further evaporation of water, arch-shaped segregation areas grow, and then merge with their neighbors [Fig.~\ref{fgr:nucleation}(a2-a4)]. Eventually, they fully occupy the entire droplet [Fig.~\ref{fgr:nucleation}(a5)]. Fig.~\ref{fgr:nucleation}(b) shows the temporal evolution of the azimuthal undulations W($\theta,t$) of the growing arches, namely the radial distance between the tip of the arches and the contact line of the droplet. The total number $N_s$ and projected area $A_s$ of the arches are displayed in the inset of Fig.~\ref{fgr:nucleation}(c). 

We interpret that the growth of liquid-liquid-phase-segregation patterns is controlled by the depleted water concentration due to evaporation. The mass reduction rate of the mixture phase is proportional to the evaporative flux rate of water, i.e., $\textrm{d}m_m/\textrm{d}t \sim -R_m D_{w,air}\Delta c$~\citep{sobac2014,li_jfm_2020}, where $R_m$ the distance between the center of the droplet to the front of the arches [see Fig.~\ref{fgr:nucleation}(a3)], $D_{w,air}$ the diffusion coefficient of water vapour in air, and $\Delta c$ the vapour concentration difference between the droplet surface and the surroundings.
The concentration dependence of the mixture liquid density $\rho_m$ is neglected here. With the geometrical relations among the volume $V_m$, the projected area $A_m$ and the radius $R_m$ of the mixture phase, i.e., $V_m \sim A_m^{3/2}$ and $R_m \sim A_m^{1/2}$, the change of the droplet mass is 

\begin{equation}
\frac{\textrm{d}m_m}{\textrm{d}t} = \frac{\textrm{d}(\rho_mV_m)}{\textrm{d}t} = \frac{\textrm{d}\rho_m}{\textrm{d}t}V_m + \rho_m \frac{\textrm{d}V_m}{\textrm{d}t},
\label{eq:tot_change_d_m}
\end{equation}
in which $\rho_m = (\rho_wV_w+\rho_HV_H)/(V_w+V_H) \approx \textrm{constant}$, with the density of water $\rho_w = 997$ kg/m$^3$ and the density of 1,2-hexanediol $\rho_H = 951$ kg/m$^3$ at the room temperature.  So we obtain

\begin{equation}
\frac{\textrm{d}m_m}{\textrm{d}t} \approx \rho_m \frac{\textrm{d}V_m}{\textrm{d}t} \sim \rho_m\frac{\textrm{d}(A_m^{3/2})}{\textrm{d}t},
\label{eq:change_d_m}
\end{equation}
and the mass depletion by evaporation 
\begin{equation}
\frac{\textrm{d}m_m}{\textrm{d}t} \sim -R_m D_{w,air}\Delta c \sim -A_m^{1/2}D_{w,air}\Delta c. 
\label{eq:depletion_m_d}
\end{equation}
 From equating Eq.~(\ref{eq:change_d_m}) and Eq.~(\ref{eq:depletion_m_d}), we obtain $\rho_m\textrm{d}(A_m^{3/2})/\textrm{d}t \sim -A_m^{1/2}D_{w,air}\Delta c$, which yields $\textrm{d}A_m \sim -D_{w,air}\Delta c/\rho_m\cdot\textrm{d}t$, i.e., linear shrinkage in time for the projected area of the mixture phase $A_m = A_t -D_{w,air}\Delta c/\rho_m \cdot t$. Here $A_t$ is the total initial footprint area and $D_{w,air}$ and $\Delta c$ were assumed to remain nearly constant. The segregated area $A_s$ then is $A_s~=~A_t-A_m$. Rescaling with the lifetime of the droplet $t_f$ and the time until the emergence of the first segregation $t_s$, which are all constant for each case, yields
\begin{equation}
\frac{A_s}{A_t} \sim \frac{t-t_s}{t_f-t_s},
\label{eq:dadt}
\end{equation}
i.e., linear growth of the segregated area. Eq.~(\ref{eq:dadt}) is verified in Fig.~\ref{fgr:nucleation}(c) by displaying the temporal evolution of the area of the segregated phase, compensated by the total footprint area $A_s/A_t$ in logarithmic scale. The observed linear growth indeed supports that the growth rate of the segregated phase patterns is controlled by the evaporation rate of water.

\subsection{Local flow structure within segregative patterns}

Following the understanding of the growth dynamics of the phase segregation of 1,2-hexanediol, we further conducted a direct measurement of the local flow structure within the segregative patterns, in order to better understand the observed radial flow reversal at Stage~II.

\begin{figure}
\centering
\includegraphics[width=1\textwidth]{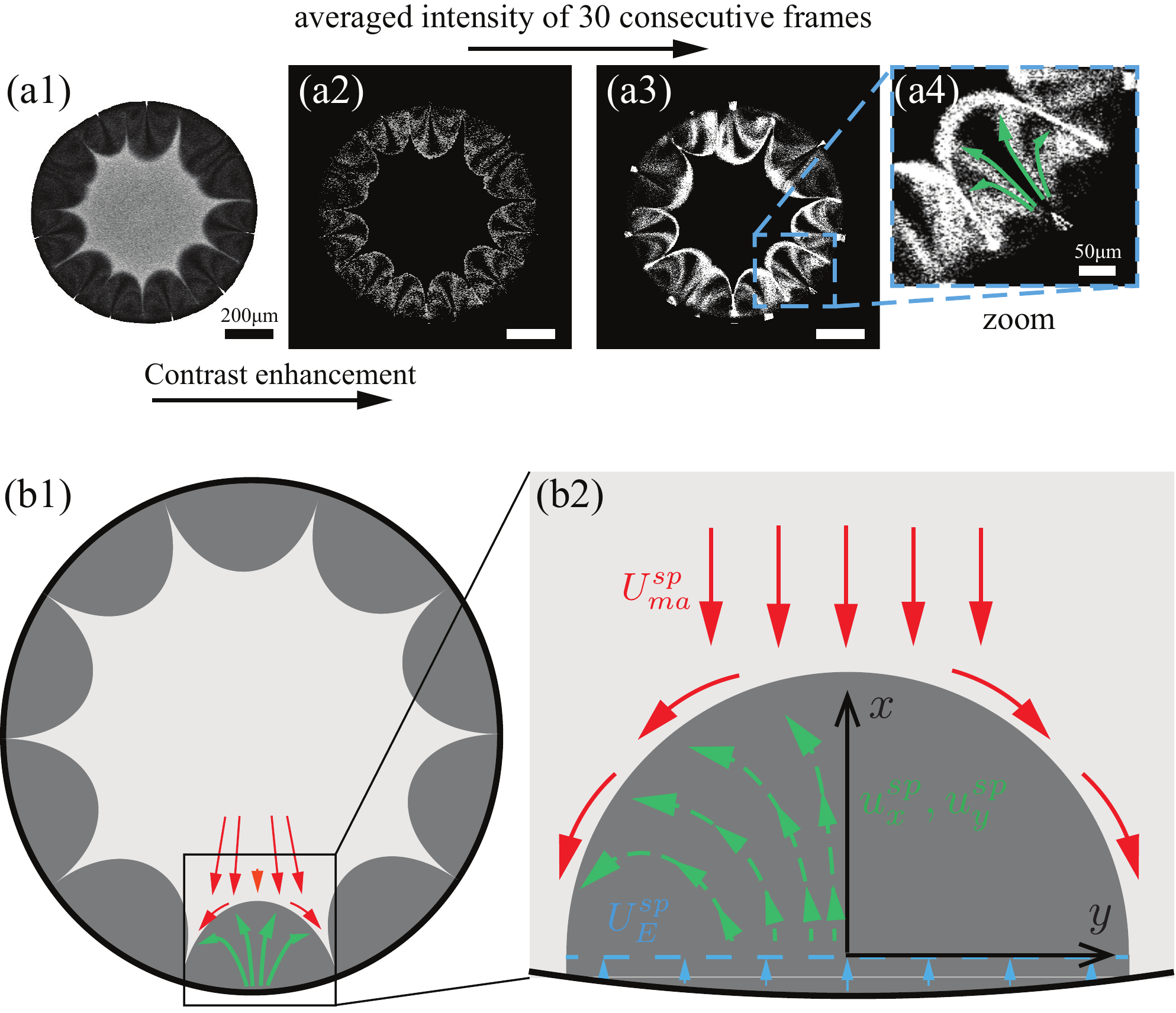}
\caption{Flow structure within the segregated phase patterns. (a1) The fluorescent signal at the focal plane near the substrate. (a2) The fluorescent signal of Dextran within segregative phase patterns. (a3) Contrast-enhanced image by averaging the intensity of 30 consecutive frames. (a4) A zoom-in image of the signal within one of the patterns. (b1-b2) Schematics of the flow structure within a segregative pattern.}
\label{fgr:velocity_cell}
\end{figure}

We utilized the water-soluble dye Dextran as ``tracer" to realize the flow visualization in those regions. The dye is originally dissolved in the mixture system but then separates from the droplet together with 1,2-hexanediol at the contact line due to the preferential evaporation of water there. Some of the separated Dextran attaches on the solid surface, causing pinning of the contact line. Some other Dextran is transported into the 1,2-hexanediol phase by the local flow [see Fig.~\ref{fgr:velocity_cell}(a)]. Because of its insolubility in 1,2-hexanediol, Dextran aggregates as small clusters in the segregated phase, which then play the role as tracer particles [see Fig.~\ref{fgr:velocity_cell}(a)]. By extracting the fluorescent signal of Dextran, we can visualize the trajectories of these tracers, which are indicative of the flow in the arches. We average the intensity of 30 consecutive images to qualitatively illustrate the flow structure within the arches, as shown in Fig.~\ref{fgr:velocity_cell}(a2-a4). 

From that figure we see that the flow structure inside the arch-like patterns consists of two counter-rotating convective rolls originating from the periphery of the binary droplet, which is indicated by the green arrows in Fig.~\ref{fgr:velocity_cell}(a4). It is reminiscent of the convective rolls inside the arch-shaped patterns observed in an evaporative ethanol-water binary system confined in a Hele-Shaw cell, which is driven by a Marangoni instability caused by the gradient of ethanol at the interface~\citep{Linde1964,ricardo2021}. However, in our case, the generation of the convective rolls follows a different route. It is schematically shown in Fig.~\ref{fgr:velocity_cell}(b1-b2): the expansion of the arches induces the flow originating from the edge towards the tip of the arches; then it encounters the outward radial flow outside the arches (indicated by red and orange arrows)[Fig.~\ref{fgr:velocity_cell}(b1)], and recirculates backwards to the edge. Note that the flow differs both from capillary flows caused by a pinned contact line and from Marangoni flows driven by surface tension gradients; instead, remarkably, it is driven by the growth of the segregative patterns.

\begin{figure}
\centering
\includegraphics[width=1\textwidth]{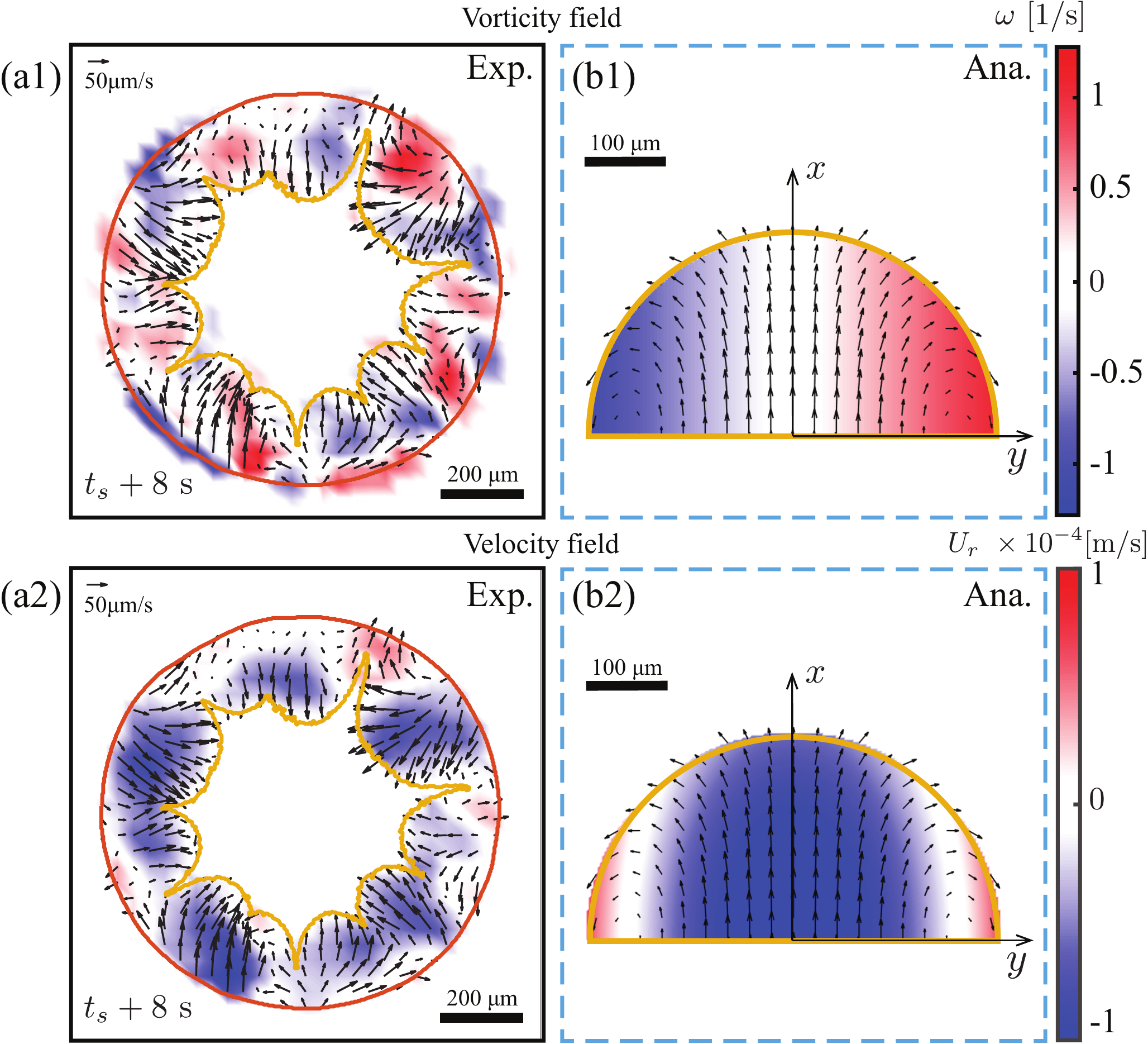}
\caption{(a1-a2) Experimental results of the flow field within the arch-shaped patterns. The colormap displays the vorticity (a1) and radial velocity (a2). (b1-b2) Analytical prediction of the flow field within one segregative pattern.}
\label{fgr:the_ana}
\end{figure}

To quantitatively confirm that the radial flow is mainly driven by the growing phase segregation, we first performed $\mu$PIV within the segregative patterns by utilizing the demixing Dextran as $\mu$PIV ``tracers". A typical snapshot of the flow field within the segregative patterns is shown in Fig.~\ref{fgr:the_ana}(a1-a2). A proper $\mu$PIV analysis is only possible after the segregation reaches a certain size, so that the area of each pattern is 3-4 times larger than the interrogation window size. 

We argue that the flow structure within every individual arch-shaped pattern is induced by the expansion of the segregation and its interaction with the surrounding flow. Due to the relative flatness of the droplet ($h/R \ll 1$), we consider the system in a two-dimensional (2D) planar geometry, namely the 2D incompressible flow is predominant in our system. To obtain an analytically solvable model, a few assumptions are taken to simplify the system: (i) Every individual segregative pattern is simplified as a half circular domain by neglecting the curvature at the contact line. (ii) For the small timescale in which the patterns remain considerably the same shape, the mass conservation holds in the half circular domain which is pushed towards the interior of the droplet by the growing segregation at the contact line region with the velocity $U^{sp}_E \approx 50~\mu$m/s. (iii) The coalescence between neighbouring patterns is not considered. We then obtain the theoretical flow velocity $u^{sp}_x, u^{sp}_y$ within the segregative region ($x\leq R_{sp},|y|\leq R_{sp}, x^2+y^2 \leq R_{sp}^2$) in a Cartesian coordinates $(x,y)$ in the laboratory frame of reference as shown in Fig.~\ref{fgr:velocity_cell}(b2) (see Supplementary Materials for more details):

\begin{equation}
u^{sp}_x = (1-\frac{x^2+3y^2}{R_{sp}^2})(U^{sp}_{ma}+U^{sp}_E)+U^{sp}_E, 
\label{eq:u_x}
\end{equation}

\begin{equation}
u^{sp}_y = \frac{2xy}{R^2_{sp}}(U^{sp}_{ma}+U^{sp}_E), 
\label{eq:u_y}
\end{equation}
where $U^{sp}_{ma} \approx 20~\mu$m/s is the velocity of the convective flow induced by Marangoni effect and $R_{sp}$ the radius of the segregative pattern.

Fig.~\ref{fgr:the_ana} displays the comparison between the experimental results and the analytical predictions by Eqs~(\ref{eq:u_x}\&\ref{eq:u_y}), which shows a good agreement of vorticity and velocity fields within the segregative patterns. It indicates that the inward flow structure (flow reversal) at the contact line region is induced by the expansion of the segregative patterns, and the two counter-rotating circulations are caused by the interaction between the expansion of the segregation and its surrounding outward flow.

\section{Stage III: Decay of flow intensity}\label{stageiii}

When the segregated 1,2-hexanediol phase evolves to an almost full occupation of the entire droplet [see Fig.~\ref{fgr:nucleation}(a5)], the evaporation process enters its final Stage III, which is characterized by the decay of the flow intensity. In this section, we quantitatively study the decay of the flow intensity by the spatially-averaged vorticity. A similar decrease of the magnitude of vorticity has been observed within the flow transition in an evaporating ethanol-water binary droplet~\citep{christy2011,bennacer_sefiane_2014,diddens2017evaporating}. The transition in that flow occurs at the period when ethanol is almost depleted by evaporation. 

\begin{figure}
\centering
\includegraphics[width=1\textwidth]{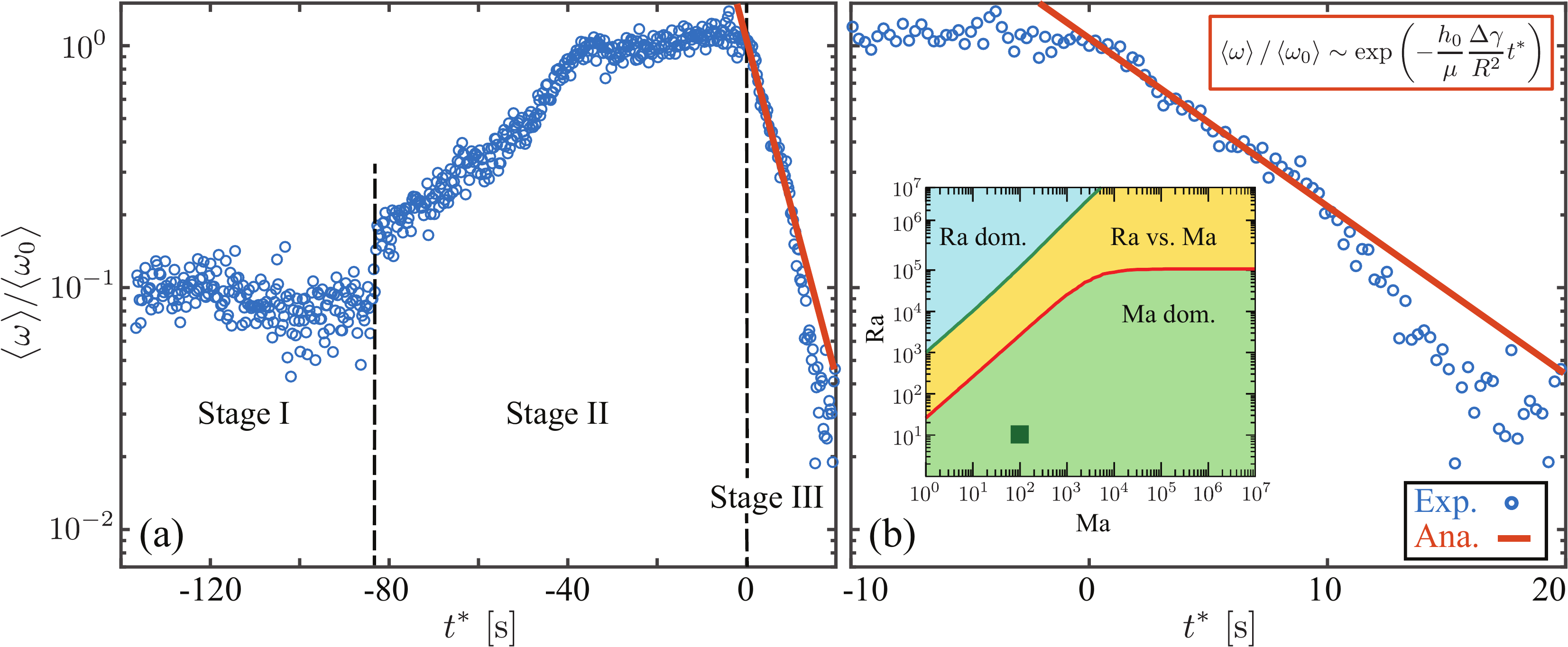}
\caption{Temporal evolution of the normalized vorticity within an evaporating 1,2-hexanediol-water droplet at the focal plane near the substrate. The blue dots represent the experimental measurement, and the red line shows the analytical prediction from Eq.~(\ref{eq:omega-t}) for the late phase, with $h_0 \approx 80~\mu$m, dynamics viscosity $\mu = 80\times 10^{-3}$~kg~m$^{-1}$~s$^{-1}$. The time $t^* = 0$ s indicates the start of the vortex decay. (a) The measured temporal spatially-averaged vorticity $\left< \omega \right>$ normalized by the value $\left< \omega_0 \right>$ at the beginning of the decay ($t^* = 0$) for the whole drying process. (b) The same measured dataset in Stage III. The inset shows the prediction by ~\cite{did21a}. The dark green bullet at $Ma = 10^2, Ra = 10^1$ indicates a typical experimental data point of this work, which clearly lies in the Marangoni-flow-dominant regime, as all data points of this work.}
\label{fgr:vortex-decay}
\end{figure}

In our 1,2-hexanediol-water droplet, we notice that the start of the decay of the mean vorticity $\left< w(t)\right>$ corresponds to the moment when the convective cells disappear. After that, the contact area still slightly shrinks, reflecting that the contact angle recovers from the dynamic receding angle to the final equilibrium angle~\citep{Li2018}. We argue that at this stage, there is still some water entrapped in the residue of the droplet, which causes a weak surface tension gradient at the liquid-air interface which drive some solute Marangoni flow. We estimate the Marangoni number $Ma = (\Delta \gamma R)/(\mu D_{w,h}) \approx 10^{2}$, with the diffusion coefficient $D_{w,h} \approx 10^{-10}$ m$^2$/s for water in 1,2-hexanediol, the dynamic viscosity $\mu \approx 80 \times 10^{-3}$ kg m$^{-1}$ s$^{-1}$, and the surface tension difference $\Delta \gamma \approx 10~\mu$N/m. $\Delta \gamma$ is estimated from the numerical simulation in our previous work~\citep{Li2018}, which shows that in the late stage of evaporation, the variance of mass concentration of 1,2-hexanediol along the droplet surface has the order of $\sim 1 \%$. It corresponds to a surface tension variance $\Delta \gamma \sim 10~\mu$N/m~\citep{Romero2007a}. Furthermore, the Rayleigh number in our case $Ra = (\Delta\rho R^3g)/(\mu D_{w,h}) \approx 10$, with density difference $\Delta\rho \sim 1$~kg m$^{-3}$, and gravitational acceleration $g = 9.8$~m s$^{-2}$. Therefore, given these estimation of $Ma$ and $Ra$, the Marangoni convection is stronger than the Rayleigh convection in this life stage~\citep{did21a}, see in the inset of Fig.~\ref{fgr:vortex-decay}(b).

The out-of-equilibrium state is generated by the non-uniform concentration distribution. The dynamic mixing can be described by the general advection-diffusion equation $\partial_tC=\nabla\cdot(D\nabla C)-\nabla\cdot(\vec{u}C) + S$, where $S$ describes the source term. In this late stage of the evaporation process, we neglect the source term $S$, take the diffusion coefficient $D$ as constant, and describe the velocity field as an incompressible flow which has zero divergence (i.e., $\nabla\cdot\vec{u} = 0$). Then the formula simplifies to 
$\partial_tC=\nabla\cdot(D\nabla C)-\vec{u}\cdot \nabla C$. With the typical velocity $U \approx 10^{-5}$ m/s, the P\'{e}clet number can be estimated by $Pe = UR/D_{w,h} \approx 10$, which reflects the dominating role of advection as compared to diffusion in this case. In order to capture the scales of both sides of the equation at leading order, we further neglect the diffusion term, hence

\begin{equation}
\partial_t C \sim -\vec{u}\cdot \nabla C.
\label{eq:con-diff}
\end{equation}

The typical flow velocity $\vec{u}$ is characterized by the velocity at the liquid-air interface driven by the surface tension gradient. It is expected to have the magnitude of $(h_0\Delta \gamma)/(\mu R)$~\citep{kim2016controlled} throughout the droplet, where $h_0$ is the height of the droplet. We then take the spatial derivative for both sides of Eq.~(\ref{eq:con-diff}),

\begin{equation}
\partial_t(\nabla C) \sim -\frac{h_0}{\mu}\frac{\Delta \gamma}{R}\cdot \nabla(\nabla C).
\label{eq:con-diff2}
\end{equation}

The local concentration gradient $\nabla C$ induce the surface tension gradient, which results in a local shear stress $\tau \sim \nabla \gamma \sim \nabla C\cdot\partial_c\gamma$. The vorticity perpendicular to the image plane is induced by the local stress $\omega_z \propto \tau$, which is defined as the curl of the in-plane velocity $(u_x, u_y)$, i.e., $\omega_z = |\nabla \times \vec{u}|_z = |\partial_x u_y - \partial_y u_x|$. Here we only consider the vorticity perpendicular to the image plane $\omega_z$, which is dominant over the other two components due to the low contact angle of the droplet. We hence argue that a scaling relation $\omega_z \propto \nabla C$ holds~\citep{christy2011}. By replacing $\nabla C$ by $\omega_z$, we then obtain the scaling relation

\begin{equation}
\frac{\partial\omega_z}{\partial t} \sim -\frac{h_0}{\mu}\frac{\Delta \gamma}{R}\cdot \nabla\omega_z \sim -\frac{h_0}{\mu}\frac{\Delta \gamma}{R}\frac{\omega_z}{R},
\label{eq:con-diff3}
\end{equation}
implying
\begin{equation}
\frac{d\omega_z}{\omega_z} \sim -\frac{h_0\Delta\gamma}{\mu R^2}dt.
\label{eq:con-diff33}
\end{equation}
By taking the vorticity $\omega_0$ at the beginning moment $t_d$ of the decaying as the initial condition, integration in time gives an exponential decaying of the vorticity,

\begin{equation}
\omega \sim \omega_0\textrm{exp}\left(-\frac{h_0\Delta \gamma}{\mu R^2}t^*\right),
\label{eq:omega-t}
\end{equation}
with the initial vorticity $\omega_0$ and $t^* = t-t_d$, where $t_d$ is the beginning moment of the decay.

Fig.~\ref{fgr:vortex-decay} shows good agreement between the analytical prediction and the measurement of the normalized mean vortex $\left< \omega \right>/\left< \omega_0 \right>$ during decaying. Note that beyond 10 s after the start of the decaying, the vorticity experiences a more rapid decay than predicted by our model. At this late stage of decaying, the flow velocity $U$ decreases to less than $10^{-6}$ m/s, which corresponds to $Pe < 1$. Besides the convection, the mutual diffusion also plays a significant role in the dynamical mixing to homogenize the concentration gradient. Hence the intensity of the flow decays faster than the prediction by the model which only considers the convective mixing. The solution indicates that the decaying of vorticity is mainly due to the mixing effect by the surface tension gradient, and the viscosity delays the mixing process by slowing down the convective flow. 

\section{Concluding remarks and discussions}\label{con}

In summary, we have studied the physiochemical hydrodynamics of an evaporating 1,2-hexanediol-water binary droplet on a flat substrate, in which the preferential evaporation of water leads to the segregation of 1,2-hexanediol. We quantitatively characterize the evolution of convective flow structures and identify three successive life stages by flow transitions, namely Stage I, II \& III. Stage I is defined as the period before the segregation occurs. The flow is mainly controlled by the Marangoni effect caused by the surface tension gradient induced by selective evaporation. At Stage II, a radial flow reversal originating from the contact line area is observed with the emergence of the segregation. We show that the growth rate of the segregative patterns is controlled by the depletion of water due to evaporation. Then we visualize the flow structure within the segregative arch-shaped patterns and interpret that the inward radial flow is remarkably induced by the growth of the segregative patterns. At the late Stage III of the evaporation, i.e., when the non-volatile segregative phase almost fully occupies the entire droplet, a vortex decay is observed and the distribution of the entrapped water is homogenized by the Marangoni effect, which is again caused by the surface tension gradient. In contrast, the viscosity delays the mixing effect. Generally, the flow structures undergo transitions from surface-tension-driven to segregation-driven and again to surface-tension-driven states, which clearly shows the richness of the hydrodynamics induced by the segregative behavior.

This work still requires further investigations of the various processes. The complex flow structure (e.g. the convective cells) in the entire droplet is three dimensional, but our present observations have been limited to a 2D focal plane at the bottom of the droplet only. In order to comprehensively understand the flow behavior, either an experimental or a numerical technique which is able to reconstruct a complete 3D flow field is required. 

\begin{acknowledgments} 
This work is part of an Industrial Partnership Programme (IPP) of the Netherlands Organization for Scientific Research (NWO). This research programme is co-financed by Canon Production Printing Holding B.V., University of Twente and Eindhoven University of Technology. DL gratefully acknowledges support by his ERC-Advanced Grant DDD (project number 740479).
\end{acknowledgments}

\bibliographystyle{jfm}
\bibliography{jfm-refs}

\end{document}